\documentclass[12pt]{article}

\newcommand{\q}{\underline{q}}
\newcommand{\sq}{\scriptstyle{\underline{q}}}

\newcommand{\sqf}{ \scriptstyle{\underline{q}f} }

\usepackage{graphicx}

\begin{document}

\title{
Even and odd nonlinear charge coherent states and their
nonclassical properties}
\author{{X.-M. Liu$~^{{\tt *}}$, B. Li} \\
{\small Department of Physics, Beijing Normal University,
Beijing 100875, China }\\
}
\date{}
\maketitle


\begin{center}
\begin{minipage}{120mm}
\vskip 0.15in
\baselineskip 0.374in
\begin{center}{\bf Abstract}\end{center}
\mbox{}\hspace{6mm}The (over)completeness of even and odd nonlinear charge coherent states is proved and their generation explored. They are demonstrated to be generalized entangled nonlinear coherent states. A
$D$-algebra realization of the SU$_f$(1,1) generators is given in terms of them. They
are shown to exhibit SU$_f$(1,1) squeezing and two-mode $f$-antibunching for some particular types of $f$-nonlinearity, but
neither one-mode nor two-mode $f$-squeezing.

PACS number(s): 42.50.Dv; 03.65.-w; 03.65.Ca; 03.65.Fd; 42.50.-p
\end{minipage}
\end{center}

\vskip 1mm 

{\small ${}^*$ Corresponding author.}

{\small ~~E-mail addresses: liuxm@bnu.edu.cn (X.-M. Liu).}



\newpage

\baselineskip 0.374in

\section*{1. Introduction}

\mbox{}\hspace{6mm}The coherent states introduced by Schr\"{o}dinger [1] and
Glauber [2] are eigenstates of the boson annihilation operator,
and have the very broad range of applications in the areas of physics [3$-$7].
However, in all the cases the quanta involved are uncharged. In 1976, Bhaumik et al.
[4,8,9] proposed the boson coherent states which, possessing definite charge,
are eigenstates of both the pair boson annihilation operator and the
charge operator. This kind of states are the so-called charge coherent
states. On the basis of this work, the charge coherent states for SU(2) [10],
SU(3) [11], and arbitrary compact Lie groups [12] were later developed.

The notion of charge coherent states has turned out to be very useful in various branches,
such as elementary particle physics [9,13$-$17], quantum field theory
[12,18,19], nuclear physics [20], thermodynamics [21$-$23], quantum
mechanics [24], and quantum optics [25$-$27].
Some proposals for implementations of such states in quantum optics were also put forward [25,26,28$-$33].

The charge coherent states, also called pair coherent states, are a type of
correlated two-mode states, which are not only important for studying fundamental
aspects of quantum mechanics, such as quantum nonlocality (Bell-inequality violation)
tests [34$-$36], but also useful in quantum information processing, such as quantum
entanglement [37,38] and quantum teleportation [39].

As is well known, the even and odd coherent states [40] are two
orthonormalized eigenstates of the square of the boson annihilation
operator, and attain an important position in the study of quantum optics
[41$-$43]. Motivated by this idea, one of the authors (X.-M.L.) [44] has
extended the charge coherent states to the even and odd charge coherent
states, defined as two orthonormalized eigenstates of both the square of
the pair boson annihilation operator and the charge operator.

Quantum groups [45,46], introduced as a mathematical
description of deformed Lie algebras, have provided the possibility of
extending the concept of coherent states to the case of $q$-deformations
[47$-$51]. A $q$-deformed harmonic oscillator [47,52] is represented by
the $q$-boson annihilation and creation operators, which satisfy the
quantum ($q$-deformed) Heisenberg-Weyl algebra [47,52,53], the latter being an elementary object in
quantum groups. The $q$-deformed coherent states advanced by Biedenharn
[47] are eigenstates of the $q$-boson annihilation operator. Such states
have been well investigated [48,49,54,55], and used extensively in quantum optics
and mathematical physics [51,56$-$60]. Moreover, the $q$-deformed charge
coherent states [61,62] were introduced as eigenstates of both the
pair $q$-boson annihilation operator and the charge operator.

A straightforward generalization of the $q$-deformed coherent states is to define the
even and odd $q$-deformed coherent states [63], which are two
orthonormalized eigenstates of the square of the $q$-boson annihilation
operator. On a similar route, the authors (X.-M.L. and C.Q.) [64] have extended
the $q$-deformed charge coherent states to the even and odd $q$-deformed charge coherent states,
defined as two orthonormalized eigenstates of both the square of the pair $q$-boson annihilation
operator and the charge operator.

Study of $q$-deformed harmonic oscillators has shown that these dynamical systems can be
interpreted as nonlinear oscillators with specific expotential dependence of the frequency
on vibration amplitude [65], thus inspiring the extension of the $q$-deformations to the
$f$-deformations for which the dependence of frequency on the amplitude is described by an
arbitrary function (called nonlinear function) [66,67]. A $f$-deformed oscillator (or nonlinear oscillator)
was defined by means of the $f$-deformed annihilation and creation operators, which satisfy the
$f$-deformed algebra (or nonlinear algebra) [66,68$-$73]. The nonlinear coherent states [66,70,72$-$74]
were already constructed as eigenstates of the $f$-deformed annihilation operator, and their mathematical
properties and nonclassical features discussed in detail [66,74$-$77]. A class of nonlinear coherent states
can be realized physically as the stationary states of the centre-of-mass motion of a trapped ion [74].
In addition, the nonlinear charge coherent states [78,79] were introduced as eigenstates of both the pair
$f$-deformed annihilation operator and the charge operator.

A further development of the nonlinear coherent states is performed by the
even and odd nonlinear coherent states [80,81], which are two
orthonormalized eigenstates of the square of the $f$-deformed annihilation
operator. In a parallel approach, the nonlinear charge coherent states
have been extended to the even and odd nonlinear charge coherent states [82,83], defined as two
orthonormalized eigenstates of both the square of the pair $f$-deformed annihilation
operator and the charge operator. In this paper, it is very desirable to prove a completeness relation
of the even and odd nonlinear charge coherent states, explore their generation, and study their properties in the aspects of both mathematics and quantum optics.

This paper is arranged as follows. In Section 2, a review of the even and odd nonlinear charge coherent states is presented, and the proof of their completeness relation given. Section 3 is devoted to generating such states. They are used to realize a $D$-algebra of the SU$_f(1,1)$
generators in Section 4. Their nonclassical properties, such as SU$_f(1,1)$ squeezing,
single- or two-mode $f$-squeezing, and two-mode $f$-antibunching, are studied in Section 5.
Section 6 contains a summary of the results.


\section*{\boldmath 2. Completeness of even and odd nonlinear charge coherent states}

\mbox{}\hspace{6mm}Two mutually commuting $f$-deformed oscillators are defined in terms of two pairs of independent $f$-deformed annihilation and creation operators $A_i$, $A_i^{\dagger}$ $(i=1,2)$, together with
corresponding number operators $N_i$, which are given by
\begin{eqnarray}
&&A_i=a_if(N_i)=f(N_i+1)a_i,\qquad N_i=a_i^{\dagger}a_i,\\
&&A_i^{\dagger}=f^{\dagger}(N_i)a_i^{\dagger}=a_i^{\dagger}f^{\dagger}(N_i+1),
\end{eqnarray}
where $f$ is a well behaved operator-valued function of $N_i$; and $a_i$ and $a_i^{\dagger}$ are the annihilation
and creation operators of usual linear harmonic oscillators, respectively. $f$ is a deformation function and means nonlinearity. The definition of $f$-deformed oscillators is an extension of the notion of $q$-deformed harmonic oscillators. The usual harmonic oscillator and $q$-deformed one are the two special cases of $f$-deformed oscillators with $f(N_i)=1$ and $f(N_i)=\sqrt{[N_i]/N_i}$ [84], respectively, where
\begin{equation}  \label{e3}
[x]=\frac{q^x-q^{-x}}{q-q^{-1}},
\end{equation}
with $q$ being a positive real deformation parameter.

The $f$-deformed algebra commutation relations are
\begin{equation}  \label{e4}
[{N_i},{A_i}]=-{A_i},\qquad [{N_i},A_i^{\dagger}]=A_i^{\dagger},
\end{equation}
\begin{equation}  \label{e5}
[{A_i},{A_i^{\dagger}}]=(N_i+1)f^2(N_i+1) -N_if^2(N_i),
\end{equation}
where $f$ is chosen to be real and $f^{\dagger}(N_i)=f(N_i)$. Of course, the usual Heisenberg-Weyl algebra and
$q$-deformed one are restored when $f(N_i)=1$ and $f(N_i)=\sqrt{[N_i]/N_i}$, respectively.

Let $|n\rangle_i$ $(n=0,1,2,\ldots )$ denote basis states of the Fock space for the mode $i$, where $|n\rangle_i$
are eigenstates of $N_i$ corresponding to the eigenvalues $n$. Thus, the operators $A_i$, $A_i^{\dagger}$ and $N_i$
can be realized by
\begin{eqnarray}
&&A_i=\sum_{n=0}^{\infty}\sqrt{n}f(n)|n-1\rangle_i\,{}_i\langle n|,\qquad N_i=\sum_{n=0}^{\infty}n|n\rangle_i\,{}_i\langle n|,\\
&&A_i^{\dagger}=\sum_{n=0}^{\infty}\sqrt{n}f(n)|n\rangle_i\,{}_i\langle n-1|.
\end{eqnarray}

We first briefly review the nonlinear charge coherent states and the even (odd) ones. The operators $A_1$ $(A_1^{\dagger})$ and $A_2$ $(A_2^{\dagger})$ are assigned the ``charge''
quanta 1 and $-1$, respectively. Thus the charge operator is given by
\begin{equation}
Q=N_1-N_2.  \label{e8}
\end{equation}
In view of the fact that
\begin{equation}
\lbrack Q,A_1A_2]=0,  \label{e9}
\end{equation}
the nonlinear charge coherent states are defined as eigenstates of both the pair $f$-deformed annihilation operator $A_1A_2$ $({\equiv}a_1f(N_1)a_2f(N_2))$ and the charge
operator $Q$, i.e.,
\begin{equation}
A_1A_2|{\xi },\q,f{\rangle }={\xi }|{\xi },\q,f{\rangle }, \qquad
Q|{\xi },\q,f{\rangle }=\q |{\xi },\q,f{\rangle },  \label{e10}
\end{equation}
where $\xi $ is a complex number and $\q$ is the charge number, which is a fixed integer.
Suppose the function $f(n)$ has no zeroes at positive integers. The $q$-deformed harmonic oscillator
is just in this case. With the help of the two-mode Fock space's completeness relation
\begin{equation}
\sum\limits_{m=0}^\infty \sum\limits_{n=0}^\infty |m,n{\rangle }{\langle }%
m,n|=I, \qquad |m,n{\rangle } \equiv |m{\rangle }_1|n{\rangle }_2,
\label{e11}
\end{equation}
the nonlinear charge coherent states can be expanded as
\begin{eqnarray}
&&|{\xi },\q,f{\rangle }=N_{\sqf}{ \sum_{p=\max(0,-\sq)}^\infty }\frac{{%
\xi }^{p+\min(0,\sq)}}{{[p!(p+\q)!]}^{1/2}f(p)!f(p+\q)!}|p+\q,p{\rangle }
\nonumber \\
&& \ \ \ \ \ \ \ \ \ \ =\left\{
\begin{array}{ll}
N_{\sqf}{\sum\limits_{n=0}^\infty }\frac{{\xi }^n}{{[n!(n+\sq)!]}
^{1/2}f(n)!f(n+\sq)!}|n+\q,n{\rangle }, &\q\geq 0, \\[0.2cm]
N_{\sqf}{\sum\limits_{n=0}^\infty }\frac{{\xi }^n}{{[n!(n-\sq)!]}
^{1/2}f(n)!f(n-\sq)!}|n,n-\q{\rangle }, & \q\leq 0,
\end{array}
\right.  \label{e12}
\end{eqnarray}
where the normalization factor $N_{\sqf}$ is given by
\begin{equation}
N_{\sqf}{\equiv }N_{\sqf}({|\xi |}^2)={ \left\{ \sum\limits_{n=0}^\infty
\frac{{({|\xi |}^2)}^n}{n!(n+|\q|)!{[f(n)!f(n+|\q|)!]}^2 }\right\} }^{-{1/2}},
\label{e13}
\end{equation}
and
\begin{equation}  \label{e14}
f(n)!=f(n)f(n-1){\cdots}f(1),\qquad f(0)!=1.
\end{equation}
It is noted that the above nonlinear charge coherent states are somewhat different from those defined
in Ref. [78] ([79]), the latter being eigenstates of both the operator $a_1f_1(N_1)a_2f_2(N_2)$ $(a_1a_2g(N_1,N_2))$
and $Q$, where $f_i$ is a function of $N_i$; $g$ is that of $N_1$ and $N_2$. The case of $f_1=f_2=f$
$(g(N_1,N_2)=f(N_1)f(N_2))$ coincides with the results given above.

The even and odd nonlinear charge coherent states are defined as two orthonormalized eigenstates of both the square ${(A_1A_2)}^2$ and the operator $Q$, i.e.,
\begin{equation}
{(A_1A_2)}^2|{\xi },\q,f{\rangle }_{e(o)}={\xi }^2|{\xi },\q,f{\rangle }%
_{e(o)}, \hspace{0.2cm} Q|{\xi },\q,f{\rangle }_{e(o)}=\q|{\xi },\q,f{%
\rangle }_{e(o)}, \hspace{0.2cm} {}_e{\langle }\xi ,\q,f|\xi ,\q,f{\rangle }%
_o=0.  \label{e15}
\end{equation}
The solutions to Eq. (15) are
\begin{eqnarray}
&&|{\xi },\q,f{\rangle }_e=N_{\sqf}^e{ \sum_{p=\max(0,-\q/2%
)}^\infty }\frac{{\xi }^{2p+\min(0,\sq)}}{{[(2p)!(2p+\q)!]}^{1/2}f(2p)!f(2p+\q)!}
|2p+\q,2p{\rangle } \nonumber \\
&& \ \ \ \ \ \ \ \ \ \ \ =\left\{
\begin{array}{ll}
N_{\sqf}^e{\sum\limits_{n=0}^\infty }\frac{{\xi }^{2n}}{{[(2n)!(2n+\sq%
)!]}^{1/2}f(2n)!f(2n+\sq)!}|2n+\q,2n{\rangle }, & \q\geq 0, \\[0.2cm]
N_{\sqf}^e{\sum\limits_{n=0}^\infty }\frac{{\xi }^{2n}}{{[(2n)!(2n-\sq%
)!]}^{1/2}f(2n)!f(2n-\sq)!}|2n,2n-\q{\rangle }, & \q\leq 0,
\end{array}
\right.  \label{e16}
\end{eqnarray}
\begin{eqnarray}
\lefteqn{|{\xi },\q,f{\rangle }_o} \nonumber \\
&&=N_{\sqf}^o{ \sum_{p=\max(0,-\q/2%
)}^\infty }\frac{{\xi }^{2p+1+\min(0,\sq)}}{{[(2p+1)!(2p+1+\q)!]}%
^{1/2} f(2p+1)!f(2p+1+\q)!}|2p+1+\q,2p+1{\rangle } \nonumber \\
&&=\left\{
\begin{array}{ll}
N_{\sqf}^o{\sum\limits_{n=0}^\infty }\frac{{\xi }^{2n+1}}{{[(2n+1)!(2n+1+%
\sq)!]}^{1/2}f(2n+1)!f(2n+1+\sq)!}|2n+1+\q,2n+1{\rangle }, & \q\geq 0,
\\
N_{\sqf}^o{\sum\limits_{n=0}^\infty }\frac{{\xi }^{2n+1}}{{[(2n+1)!(2n+1-%
\sq)!]}^{1/2}f(2n+1)!f(2n+1-\sq)!}|2n+1,2n+1-\q{\rangle }, & \q\leq 0,
\end{array}
\right.  \label{e17}
\end{eqnarray}
where the normalization factors $N_{\sqf}^{e(o)}$ are given by
\begin{equation}
N_{\sqf}^e{\equiv }N_{\sqf}^e({|\xi |}^2)={ \left\{
\sum\limits_{n=0}^\infty
\frac{{({|\xi |}^2)}^{2n}}{(2n)!(2n+|\q|)! {[f(2n)!f(2n+|\q|)!]}^2 }%
\right\} }^{-{1/2}},  \label{e18}
\end{equation}
\begin{equation}
N_{\sqf}^o{\equiv }N_{\sqf}^o({|\xi |}^2)={ \left\{
\sum\limits_{n=0}^\infty \frac{{({|\xi |}^2)}^{2n+1}} {(2n+1)!(2n+1+|\q|)!{[f(2n+1)!f(2n+1+|\q|)!]}^2}%
\right\} }^{-{1/2}}.  \label{e19}
\end{equation}
As two special cases, for $f(n)=1$, they reduce to the usual even and odd charge coherent states constructed
by the author (X.-M.L.) [44]; for $f(n)=\sqrt{[n]/n}$, they reduce to the even and odd $q$-deformed charge coherent
states constructed by the authors (X.-M.L. and C.Q.) [64]. Note however that $|{\xi },\q,f{\rangle }_{e(o)}$ are not
eigenstates of $A_1A_2$.

We should stress that the representation (16) ((17)) of even (odd) nonlinear charge coherent states as bipartite pure
states by a biorthogonal sum is a standard form of the Schmidt decomposition [85$-$89]. Obviously, both even and odd states are entangled states since the Schmidt number for such states is infinite.

It should be remarked that the even (odd) states $|{\xi },\q,f{\rangle }_{e(o)}$ are normalizable provided $N_{\sqf}^{e(o)}$ are non-zero and finite. This means that the terms in summation for ${(N_{\sqf}^{e(o)})}^{-2}$
should be such that
\begin{equation}
\label{e20}
|\xi| < \lim_{n \rightarrow {\infty}} nf^{2}(n).
\end{equation}
If $|f(n)|$ decreases faster than $n^{-{1\over 2}}$ for large $n$, then the range of $\xi$, for which
the states $|{\xi },\q,f{\rangle }_{e(o)}$ are normalizable, is restricted to values satisfying (20) and in other cases the range of $\xi$ is unrestricted.

{}From (16) and (17), it follows that
\begin{eqnarray}
&&{}_{e(o)}{\langle }\xi ,\q,f|{\xi }^{\prime },\q^{\prime },f{\rangle }%
_{e(o)}=N_{\sqf}^{e(o)}({|\xi |}^2)N_{\sqf}^{e(o)}({|{\xi }^{\prime }|}^2){%
\left[ N_{\sqf}^{e(o)}({\xi }^{*}{\xi }^{\prime })\right] }^{-2}{\delta }_{%
\q,\q'}, \\
&&{}_e{\langle }\xi ,\q,f|{\xi }^{\prime },\q^{\prime },f{\rangle }_o=0.
\end{eqnarray}
This further indicates that the even (odd) nonlinear charge coherent states are orthogonal
with respect to the charge number $\q$ and that arbitrary even and odd nonlinear charge coherent states
are orthogonal to each other. However, these even (odd) states are nonorthogonal with respect to the parameter $\xi$.

{}For the mean values of the operators $N_1$ and $N_2$, there exists the
relation
\begin{equation}
{}_{e(o)}{\langle }\xi ,\q,f|N_1|{\xi },\q,f{\rangle }_{e(o)}=\q + {}_{e(o)}%
{\langle }\xi ,\q,f|N_2|\xi,\q,f{\rangle }_{e(o)}.  \label{e23}
\end{equation}

In terms of the even and odd nonlinear charge coherent states, the nonlinear charge coherent states
can be expanded as
\begin{equation}
|\xi ,\q,f{\rangle }=N_{\sqf}\left[ {(N_{\sqf}^e)}^{-1}|\xi ,\q,f{%
\rangle }_e+{(N_{\sqf}^o)}^{-1}|\xi ,\q,f{\rangle }_o\right], \label{e24}
\end{equation}
where the normalization factors are such that
\begin{equation}
N_{\sqf}^{-2}={(N_{\sqf}^e)}^{-2}+{(N_{\sqf}^o)}^{-2}.  \label{e25}
\end{equation}

To make up a completeness relation of the even and odd nonlinear charge coherent states, we introduce another
type of even and odd nonlinear charge coherent states $|{\xi },\q,f{\rangle }{\rangle }_{e(o)}$, which are two orthonormalized eigenstates of both the square ${(\tilde{A}_1\tilde{A}_2)}^2$ and the operator $Q$, where
\begin{equation}
\tilde{A}_i=a_i\frac{1}{f(N_i)}.  \label{e26}
\end{equation}
It is evident that the representations of $|{\xi },\q,f{\rangle }{\rangle }_{e(o)}$ in the two-mode Fock space are the same as $|{\xi },\q,f{\rangle }_{e(o)}$ given in Eqs. (16) and (17) except for replacing $f(n)$ by $\frac{1}{f(n)}$.

We now prove that the even and odd nonlinear charge coherent states form an (over)complete set, that is to say
\begin{eqnarray}
&&\sum\limits_{\sq={-\infty }}^\infty \int \frac{d^2{\xi }}\pi 2{|\xi|}^{|\sq|}K_{\sq}(2|\xi|)
\left[ {(N_{\sqf}^eN_{\sq \frac{1}{f} }^e)}^{-1}|\xi ,\q,f\rangle _e\,{}_e{%
\langle }{\langle}\xi ,\q,f|+{(N_{\sqf}^oN_{\sq \frac{1}{f} }^o)}^{-1}|\xi ,\q,f\rangle _o\,{}_o{\langle }{\langle }%
\xi ,\q,f|\right] \nonumber \\
&&\equiv \sum\limits_{\sq={-\infty }}^\infty I_{\sq}=I,
\end{eqnarray}
where $K_{\sq}(Z)$ is the second kind of modified Bessel function of order $\q$ [90],
and $N_{\sq \frac{1}{f} }^{e(o)}$ is a normalization factor of $|{\xi },\q,f{\rangle }{\rangle }_{e(o)}$.

In fact, for $\q\geq 0$, we have
\begin{eqnarray}
I_{\q} &=&\int \frac{d^2{\xi }}\pi 2{|\xi|}^{\sq}K_{\sq}(2|\xi|)
\sum\limits_{j=0}^1\sum\limits_{n,m}^{} {\xi }^{2n+j}{{\xi }^{*}}^{2m+j} \nonumber \\
&& \mbox{} {\times}
\frac{f(2m+j)!f(2m+j+\q)! |2n+j+\q,2n+j{\rangle }{\langle }2m+j+\q,2m+j|}{{ \left[(2n+j)!(2n+j+%
\q)!(2m+j)!(2m+j+\q)!\right] }^{1/2} f(2n+j)!f(2n+j+\q)!}  \nonumber \\
&=&\int\limits_0^\infty \frac{d|\xi |}\pi 2 {|\xi |}^{\sq+1} K_{\sq}(2|\xi|)
\sum\limits_{j=0}^1\sum\limits_{n,m}^{}{|\xi |}^{2(n+m+j)}\int\limits_{-\pi
}^\pi d{\theta }e^{2{\rm i}(n-m){\theta }}  \nonumber \\
&&\mbox{} {\times }
\frac{f(2m+j)!f(2m+j+\q)! |2n+j+\q,2n+j{\rangle }{\langle }2m+j+\q,2m+j|}{{ \left[(2n+j)!(2n+j+%
\q)!(2m+j)!(2m+j+\q)!\right] }^{1/2} f(2n+j)!f(2n+j+\q)!}  \nonumber \\
&=&\sum\limits_{j=0}^1\sum\limits_{n=0}^\infty \frac{4|2n+j+\q,2n+j{\rangle }{\langle }2n+j+\q,2n+j|} {(2n+j)!(2n+j+\q)!}
\int\limits_0^\infty {d|\xi |}{|\xi |}^{4n+2j+\q+1}K_{\sq}(2|\xi|)
\nonumber \\
&=&\sum\limits_{j=0}^1 I_{\q}^j \nonumber \\
&=&\sum\limits_{n=0}^\infty |n+\q,n{\rangle }{\langle }n+\q,n|,
\end{eqnarray}
where
\begin{equation}
I_{\q}^j=\sum\limits_{n=0}^\infty |2n+j+\q,2n+j{\rangle }{\langle }2n+j+\q,2n+j|, \quad j=0,1.
\label{e29}
\end{equation}
Similarly, for $\q\leq 0$, we get
\begin{eqnarray}
&&I_{\q} =\sum\limits_{j=0}^1 I_{\q}^j \nonumber \\
&&\ \ \ =\sum\limits_{n=0}^\infty |n,n-\q{\rangle }{\langle }n,n-\q|,
\end{eqnarray}
where
\begin{equation}
I_{\q}^j=\sum\limits_{n=0}^\infty |2n+j,2n+j-\q{\rangle }{\langle }2n+j,2n+j-\q|, \quad j=0,1.
\label{e31}
\end{equation}
Consequently, we derive
\begin{eqnarray}
&&\sum\limits_{\sq=-\infty }^\infty I_{\sq} =\sum\limits_{n=0}^\infty
\left( \sum\limits_{\sq=-\infty }^{-1}|n,n-\q{\rangle }{\langle
}n,n-\q|+\sum\limits_{\sq=0}^\infty |n+\q,n{\rangle }{\langle }n+\q%
,n|\right)  \nonumber \\
&& \ \ \ \ \ \ \ \ \ \ =\sum\limits_{m=0}^\infty \sum\limits_{n=0}^\infty
|m,n{\rangle }{\langle } m,n|=I.
\end{eqnarray}
Hence, the even and odd nonlinear charge coherent states are qualified to
make up an overcomplete representation. It should be mentioned that $I_{\sq}$
represents the resolution of unity in the subspace where $Q=\q$; $I_{\q}^{e(o)}({\equiv}I_{\q}^{0(1)})$
represents that in the even (odd) subspace where $Q=\q$ and satisfies
\begin{equation}
I_{\q}^j I_{\q^{\prime }}^{j^{\prime}}=I_{\q}^j{\delta }_{j,j'}{\delta }_{\q,\q'}.
\label{e33}
\end{equation}

For Eq. (27), its conjugate is
\begin{eqnarray}
&&\sum\limits_{\sq={-\infty }}^\infty \int \frac{d^2{\xi }}\pi 2{|\xi|}^{|\sq|}K_{\sq}(2|\xi|)
\left[ {(N_{\sqf}^eN_{\sq \frac{1}{f} }^e)}^{-1}|\xi ,\q,f\rangle\rangle _e\,{}_e{%
\langle }\xi ,\q,f|+{(N_{\sqf}^oN_{\sq \frac{1}{f} }^o)}^{-1}|\xi ,\q,f\rangle\rangle _o\,{}_o{\langle }%
\xi ,\q,f|\right] \nonumber \\
&&\equiv \sum\limits_{\sq={-\infty }}^\infty I_{\sq}=I.
\end{eqnarray}
Note that in (27) the ket and bra are not mutually Hermite conjugate.


\section*{\boldmath 3. Generation of even and odd nonlinear charge coherent states}

\mbox{}\hspace{6mm}From (12), (16) and (17), it follows that [82,83]
\begin{eqnarray}
&&|\xi ,\q,f\rangle_e ={\frac 12}\frac{N_{\sqf}^e}{N_{\sqf}}(|\xi ,\q,f\rangle +|-\xi
,\q,f{\rangle }), \\
&&|\xi ,\q,f{\rangle }_o ={\frac 12}\frac{N_{\sqf}^o}{N_{\sqf}}(|\xi ,\q,f\rangle -|-\xi
,\q,f{\rangle }).
\end{eqnarray}
This shows that the even (odd) nonlinear charge coherent states can be
obtained by the symmetric (antisymmetric) combination of nonlinear
charge coherent states as the charge is conserved. This is similar to the case
of even (odd) nonlinear coherent states, which are combinations of nonlinear
coherent states, namely,
\begin{eqnarray}
&&|\xi ,f{\rangle }_e ={\frac 12}\frac{N_f^e}N(|\xi ,f{\rangle }+|-\xi ,f{\rangle }%
), \\
&&|\xi ,f{\rangle }_o ={\frac 12}\frac{N_f^o}N(|\xi ,f{\rangle }-|-\xi ,f{\rangle }%
),
\end{eqnarray}
where
\begin{equation}
\label{e39}
|\xi ,f{\rangle }=N_f\sum\limits_{n=0}^\infty \frac{{\xi }^n}{\sqrt{n!}f(n)!}|n{%
\rangle },
\end{equation}
\begin{eqnarray}
&&N_{f}{\equiv }N_{f}({|\xi |}^2)={ \left\{ \sum\limits_{n=0}^\infty
\frac{{({|\xi |}^2)}^n}{n!{[f(n)!]}^2 }\right\} }^{-{1/2}}, \\
&&N_{f}^e{\equiv }N_{f}^e({|\xi |}^2)={ \left\{
\sum\limits_{n=0}^\infty
\frac{{({|\xi |}^2)}^{2n}}{(2n)!{[f(2n)!]}^2 }\right\} }^{-{1/2}}, \\
&&N_{f}^o{\equiv }N_{f}^o({|\xi |}^2)={ \left\{
\sum\limits_{n=0}^\infty
\frac{{({|\xi |}^2)}^{2n+1}} {(2n+1)!{[f(2n+1)!]}^2}\right\} }^{-{1/2}}.
\end{eqnarray}

Here, we find that the even (odd) nonlinear charge coherent states can also be generated from
the states $(37)-(39)$ according to the following expression
\begin{eqnarray}
\lefteqn{|\xi ,\q,f{\rangle }_{e(o)}} \nonumber \\
&&=\left\{
\begin{array}{ll}
N_{\sqf}^{e(o)} N_{f}^{-1} ({|{\xi }_1|}^2) {\left[ N_f^{e(o)}({|{\xi }_2|}%
^2)\right] }^{-1}{\xi _1}^{-\sq}\int\limits_{-\pi }^\pi \frac{d{\alpha }}{%
2\pi }e^{+{\rm i}\sq\alpha }|e^{-{\rm i}\alpha }{\xi }_1 ,f{\rangle }{\otimes
}|e^{{\rm i}\alpha }{%
\xi }_2 ,f{\rangle }_{e(o)}, & \q\geq 0, \\[0.2cm]
N_{\sqf}^{e(o)} N_{f}^{-1} ({|{\xi }_1|}^2) {\left[ N_f^{e(o)}({|{\xi }_2|}%
^2)\right] }^{-1}{\xi _1}^{+\sq}\int\limits_{-\pi }^\pi \frac{d{\alpha }}{%
2\pi }e^{-{\rm i}\sq\alpha }|e^{{\rm i}\alpha }{\xi }_2 ,f{\rangle }_{e(o)}{\otimes }%
|e^{-{\rm i}\alpha }{\xi }_1 ,f{\rangle }, & \q\leq 0,
\end{array}
\right.  \label{e43}
\end{eqnarray}
where $\xi ={\xi }_1{\xi }_2$. Such a representation is very useful since
the properties of nonlinear coherent states and even (odd) nonlinear
coherent states can now be employed in a study of the properties of even
(odd) nonlinear charge coherent states. The expression for the latter given in (43)
has a very simple group-theoretical interpretation: in (43) one suitably averages over the
U(1)-group (caused by the charge operator $Q$) action on the product of
nonlinear coherent states and even (odd) nonlinear coherent states, which
then projects out the $Q=\q$ charge subspace contribution.

It is easy to see that in the two special cases of $f(n)=1$ and $f(n)=\sqrt{[n]/n}$, the above discussion
gives back the corresponding results for the usual even (odd) charge coherent states obtained in Ref. [44] and the even (odd) $q$-deformed charge coherent states in Ref. [64], respectively.

On the other hand, from the entangled nonorthogonal state point of view, the even (odd) nonlinear charge coherent states given in (43) are represented as a continuous entangled sum of nonlinear coherent states and even (odd) nonlinear coherent states. Therefore, following the definition of entangled coherent states [91$-$93], we call
them generalized entangled nonlinear coherent states.



\section*{\boldmath 4. $D$-algebra realization of SU$_f(1,1)$ generators}

\mbox{}\hspace{6mm}As is well known, the coherent state $D$-algebra [6,94] is a
mapping of quantum observables onto a differential form that acts on the parameter space
of coherent states, and has a beautiful application in the reformulation of the entire laser
theory in terms of $C$-number differential equations [95]. We shall construct the
$D$-algebra realization of the $f$-deformed SU$_f(1,1)$ generators corresponding to
the unnormalized even and odd nonlinear charge coherent states, defined by
\begin{equation}
||\q{\rangle }_{e(o)}{\equiv }||\xi ,\q,f{\rangle }_{e(o)}={\left[ N_{\sqf}^{e(o)}\right]
}^{-1}|\xi ,\q,f{\rangle }_{e(o)}.  \label{e44}
\end{equation}

Let $||\q{\rangle }$ denote a column vector composed of $||\q{\rangle }_e$ and $||\q{\rangle }_o$, i.e.,
\begin{equation}
||\q{\rangle }{\equiv }
\left[
\begin{array}{l}
||\q{\rangle }_e \\
||\q{\rangle }_o
\end{array}
\right].
\label{e45}
\end{equation}
The action of the operators $A_i$, $A_i^{\dagger}$, $\tilde{A}_i$,  $\tilde{A}_i^{\dagger}$ and $N_i$ on this
column vector can be written in the matrix form of differential operators,
\begin{equation}
\begin{array}{ll}
{\rm Positive\ } Q & {\rm Negative\ } Q \\[0.4cm]
A_1
||\q{\rangle }
=||\q-1{\rangle } & A_1
||\q{\rangle }={\xi }M
||\q-1{\rangle } \\[0.4cm]
A_2||\q{\rangle }
={\xi }M ||\q+1{\rangle } & A_2
||\q{\rangle }=||\q+1{\rangle } \\[0.4cm]
A_1^{\dagger}
||\q{\rangle }
={\xi }^{-\sq}f^2(\frac d{d\xi }\xi)\frac d{d\xi }{\xi}^{\sq+1}
||\q+1{\rangle } ~~~~~& A_1^{\dagger}
||\q{\rangle }= f^2(\frac d{d\xi }\xi)\frac d{d\xi }M
||\q+1{\rangle } \\[0.4cm]
A_2^{\dagger}||\q{\rangle }
=f^2(\frac d{d\xi }\xi)\frac d{d\xi }M
||\q-1{\rangle } & A_2^{\dagger}
||\q{\rangle }
= {\xi }^{\sq}f^2(\frac d{d\xi }\xi)\frac d{d\xi }{\xi}^{-\sq+1}
||\q-1{\rangle } \\[0.4cm]
N_1||\q{\rangle }
= \left({\xi }\frac d{d\xi }+\q \right)
||\q{\rangle } & N_1
||\q{\rangle }
= \xi\frac d{d\xi }
||\q{\rangle } \\[0.4cm]
N_2||\q{\rangle }
= {\xi }\frac d{d\xi }
||\q{\rangle } & N_2
||\q{\rangle }
= \left(\xi\frac d{d\xi }-\q\right)
||\q{\rangle } \\[0.4cm]
\tilde{A}_1||\q{\rangle }
={\xi}^{-\sq+1} \frac 1{ f^2(\frac d{d\xi }\xi) } {\xi}^{\sq-1}
||\q-1{\rangle } & \tilde{A}_1||\q{\rangle }
= {\xi } \frac 1{f^2(\frac d{d\xi }\xi)}M
||\q-1{\rangle } \\[0.4cm]
\tilde{A}_2||\q{\rangle }
={\xi } \frac 1{f^2(\frac d{d\xi }\xi)}M
||\q+1{\rangle } & \tilde{A}_2||\q{\rangle }
={\xi}^{\sq+1} \frac 1{ f^2(\frac d{d\xi }\xi) } {\xi}^{-\sq-1}
||\q+1{\rangle } \\[0.4cm]
\tilde{A}_1^{\dagger}||\q{\rangle }
= \left({\xi }\frac d{d\xi }+\q+1 \right)
||\q+1{\rangle } & \tilde{A}_1^{\dagger}
||\q{\rangle }
= \frac d{d\xi }M
||\q+1{\rangle } \\[0.4cm]
\tilde{A}_2^{\dagger}||\q{\rangle }
=\frac d{d\xi }M
||\q-1{\rangle } & \tilde{A}_2^{\dagger}
||\q{\rangle }
= \left(\xi\frac d{d\xi }-\q+1\right)
||\q-1{\rangle },
\end{array}
\label{e46}
\end{equation}
where
\begin{equation}
M=\left[
\begin{array}{lr}
0 & 1 \\
1 & 0
\end{array}
\right],
\label{e47}
\end{equation}
$\frac 1{f(\frac d{d\xi }\xi)}$ is the inverse of $f(\frac d{d\xi }\xi)$, and the action of $f(\frac d{d\xi }\xi)$ on
${\xi}^n$ is given by
\begin{equation}
f(\frac d{d\xi }\xi){\xi}^n=f(n+1){\xi}^n.
\label{e48}
\end{equation}
Some useful relations for differential operators are as follows:
\begin{eqnarray}
&&f(\frac d{d\xi }\xi)\frac d{d\xi }=\frac d{d\xi }f(\xi\frac d{d\xi }),~~~~~~\qquad {\xi}f(\frac d{d\xi }\xi)
=f(\xi\frac d{d\xi }){\xi},\\
&&{\xi}^{-n}f(\frac d{d\xi }\xi){\xi}^n=f(\frac d{d\xi }\xi+n),\qquad {\xi}^{-n}f(\xi\frac d{d\xi }){\xi}^{n}
=f(\xi\frac d{d\xi }+n),
\end{eqnarray}
where the action of $f(\xi\frac d{d\xi })$ is given by
\begin{equation}
f(\xi\frac d{d\xi }){\xi}^n=f(n){\xi}^n.
\label{e51}
\end{equation}
Comparing the formulae in (46) in the $q$-deformed case of $f(n)=\sqrt{[n]/n}$ with those obtained in Ref. [64]
for the even and odd $q$-deformed charge coherent states, and using (49) or (50), we have
\begin{equation}  \label{e52}
\frac{d}{d_q\xi}=[\frac d{d\xi }\xi]{\xi}^{-1}={\xi}^{-1}[\xi\frac d{d\xi }],
\end{equation}
where $d/d_q\xi$ is a $q$-differential operator [49,54,96], defined by
\begin{equation}  \label{e53}
\frac{d}{d_q\xi}f(\xi)=\frac{f(q\xi)-f(q^{-1}\xi)}{q\xi-q^{-1}\xi }.
\end{equation}
Thus, we get an important result that the $q$-differential operator can be realized by the standard differential operator according to the expression (52). It should be noticed that the operator ${\xi}^{-1}[\xi\frac d{d\xi }]$
in (52) was already introduced by Solomon [72].

It is easy to check that in the two special cases of $f(n)=1$ and $f(n)=\sqrt{[n]/n}$, the above discussion
gives back that carried out in Refs. [44,64] for the usual even (odd) charge coherent states and the even (odd)
$q$-deformed ones, respectively.

Let us define the $f$-deformed SU$_f(1,1)$ algebra, which consists of three generators $K_0$, $K_{+}$ and $K_{-}$, where
\begin{equation}
K_{-}=A_1A_2, \qquad K_{+}=\tilde{A}_1^{\dagger}\tilde{A}_2^{\dagger}, \qquad K_0={\frac
12}(N_1+N_2+1),  \label{e54}
\end{equation}
the latter satisfying the commutation relations
\begin{equation}
\lbrack K_{+},K_{-}]=-2K_0, \qquad [K_0,K_{\pm }]={\pm }K_{\pm },
\label{e55}
\end{equation}
while their Hermitian conjugates satisfying the dual algebra
\begin{equation}
\lbrack K_{-}^{\dagger},K_{+}^{\dagger}]=-2K_0, \qquad [K_0,K_{\mp }^{\dagger}]={\pm }K_{\mp }^{\dagger}.
\label{e56}
\end{equation}
Note that $K_0$ is Hermitian whereas $K_{+}$ and $K_{-}$ are not generally required to be Hermitian conjugate
to each other. Obviously, this algebra ia a generalization of the SU$(1,1)$ Lie algebra. When $K_{+}$ and $K_{-}$
are Hermitian conjugate to each other in the special case of $f(N_i)=1$, i.e., $K_{-}^{\dagger}=K_{+}$, the SU$_f(1,1)$ algebra contracts to the SU$(1,1)$ Lie algebra. Actually, the even and odd nonlinear charge coherent states are also eigenstates of the square of $K_{-}$.

The $D$-algebra of the SU$_f(1,1)$ generators $A$ may be defined for the action on
the ket coherent states (45) or for that on the corresponding bras as
\begin{eqnarray}
&&A||\q{\rangle }
=D^k(A)||\q{\rangle }, \label{e57}\\
&&{\langle }\q||A =D^b(A)
{\langle }\q||,
\end{eqnarray}
respectively. Using (46) and (54), we get for the former
\begin{eqnarray}
&&D^k(K_{-}) = {\xi }M, \label{e59} \\
&&D^k(K_{+})  =  \frac d{d\xi }(\xi\frac d{d\xi }+|\q|)M, \\
&&D^k(K_0)  = \frac 12 \left(2{\xi }\frac d{d\xi }+|\q|+1\right)I, \\
&&D^k(K_{-}^{\dagger})  = {\xi }^{-|\sq|}f^2(\frac d{d\xi }\xi)\frac d{d\xi }{\xi}^{|\sq|+1}
f^2(\frac d{d\xi }\xi)\frac d{d\xi }M, \\
&&D^k(K_{+}^{\dagger})  =  {\xi }^{-|\sq|+1}  \frac 1{ f^2(\frac d{d\xi }\xi) } {\xi }^{|\sq|} \frac 1{ f^2(\frac d{d\xi }\xi) }M,
\end{eqnarray}
while the latter can be obtained from the adjoint relation
\begin{equation}
D^b(A)={\left[ D^k(A^{\dagger})\right] }^{*}.  \label{e64}
\end{equation}
Thus, the $D$-algebra of the SU$_f(1,1)$ generators corresponding to the unnormalized
even and odd nonlinear charge coherent states has been realized in a differential-operator matrix form.

{}From (45), (47), (57) and (59), we clearly see that the unnormalized even
and odd nonlinear charge coherent states can be transformed into each other by the
action of the operator $A_1A_2$. Actually, $A_1A_2$ plays the role of a connecting operator between
the two kinds of states.



\section*{\boldmath 5. Nonclassical properties of even and odd nonlinear charge
coherent states}

\mbox{}\hspace{6mm}In this section, we will study some nonclassical properties of the
even and odd nonlinear charge coherent states, such as SU$_f(1,1)$ squeezing,
single- or two-mode $f$-squeezing, and two-mode $f$-antibunching.


\subsection*{\boldmath 5.1. SU$_f(1,1)$ squeezing}

\mbox{}\hspace{6mm}In analogy with the definition of SU$_q(1,1)$ squeezing [64,97], which is a $q$-deformed
analouge to SU(1,1) squeezing [44,98,99], we introduce SU$_f(1,1)$ squeezing in terms of the Hermitian
$f$-deformed quadrature operators
\begin{equation}
X_1=\frac{K_{-}^{\dagger}+K_{-}}2, \qquad X_2=\frac{{\rm i}(K_{-}^{\dagger}-K_{-})}2,  \label{e65}
\end{equation}
which satisfy the commutation relation
\begin{equation}
\lbrack X_1,X_2]=\frac{\rm i}{2}\lbrack K_{-},K_{-}^{\dagger}]=\frac{\rm i}{2}
\lbrack (N_1+1)f^2(N_1+1)(N_2+1)f^2(N_2+1) -N_1f^2(N_1)N_2f^2(N_2)]  \label{e66}
\end{equation}
and the uncertainty relation 
\begin{equation}
{\langle }{({\Delta }X_1)}^2{\rangle }{\langle }{({\Delta }X_2)}^2{\rangle }{%
\geq }\frac 1{16}{|{\langle }\lbrack K_{-},K_{-}^{\dagger}]{\rangle }|}^2.  \label{e67}
\end{equation}
A state is said to be SU$_f(1,1)$ squeezed if
\begin{equation}
{\langle }{({\Delta }X_i)}^2{\rangle }<\frac 14{|{\langle }\lbrack K_{-},K_{-}^{\dagger}]{\rangle }|}
\qquad (i=1 {\rm \ or\ }2).  \label{e68}
\end{equation}
Obviously, SU(1,1) and SU$_q(1,1)$ squeezing are the two special cases of SU$_f(1,1)$ squeezing
with $f(N_i)=1$ and $f(N_i)=\sqrt{[N_i]/N_i}$, respectively. Therefore, SU$_f(1,1)$ squeezing is a natural
extension of SU(1,1) and SU$_q(1,1)$ squeezing.

Let us now calculate the fluctuations (variances) of $X_1$ and $X_2$ with
respect to the even and odd nonlinear charge coherent states. Using (54), (57)$-$(59) and (64), we get
\begin{eqnarray}
&&{}_e{\langle }{\xi },\q,f|K_{-}^{\dagger}K_{-}|\xi ,\q,f{\rangle }_e  = {|\xi |}^2%
\, \overline{\tanh}_{\sqf}{|\xi |}^2, \\
&&{}_o{\langle }{\xi },\q,f|K_{-}^{\dagger}K_{-}|\xi ,\q,f{\rangle }_o  = {|\xi |}^2%
\, \overline{\coth}_{\sqf}{|\xi |}^2,
\end{eqnarray}
where
\begin{equation}
\overline{\tanh}_{\sqf}{|\xi |}^2 \equiv \frac{\overline{\sinh}_{\sqf}{|\xi |}^2}{%
\overline{\cosh}_{\sqf}{|\xi |}^2}, \qquad \overline{\coth}_{\sqf}{|\xi |}%
^2 \equiv \frac 1{\overline{\tanh}_{\sqf}{|\xi |}^2},  \label{e71}
\end{equation}
with
\begin{equation}
\overline{\sinh}_{\sqf}x \equiv {\left[ N_{\sqf}^o(x)\right] }^{-2}, \qquad
\overline{\cosh}_{\sqf}x \equiv {\left[ N_{\sqf}^e(x)\right] }^{-2}.  \label{e72}
\end{equation}
Furthermore,
\begin{equation}
{}_{e(o)}{\langle }K_{-}{\rangle }_{e(o)}= {}_{e(o)}{\langle }K_{-}^{\dagger}{\rangle }%
_{e(o)}=0.  \label{e73}
\end{equation}
Thus, the fluctuations are given by
\begin{eqnarray}
&&{}_e{\langle }{({\Delta }X_1)}^2{\rangle }_e = {\frac 14}{}_e{\langle }
\lbrack K_{-},K_{-}^{\dagger}]
\rangle_e+\frac 12{|\xi |}^2(\cos2\theta +\overline{\tanh}_{\sqf}{|\xi |}^2),
\label{e74}\\
&&{}_e{\langle }{({\Delta }X_2)}^2{\rangle }_e  = {\frac 14}{}_e{\langle }
\lbrack K_{-},K_{-}^{\dagger}]
\rangle _e+\frac 12{|\xi |}^2(-\cos2\theta +\overline{\tanh}_{\sqf}{|\xi |}%
^2), \\
&&{}_o{\langle }{({\Delta }X_1)}^2{\rangle }_o  = {\frac 14}{}_o{\langle }
\lbrack K_{-},K_{-}^{\dagger}]
\rangle_o+\frac 12{|\xi |}^2(\cos2\theta +\overline{\coth}_{\sqf}{|\xi |}%
^2), \\
&&{}_o{\langle }{({\Delta }X_2)}^2{\rangle }_o  = {\frac 14}{}_o{\langle }
\lbrack K_{-},K_{-}^{\dagger}]
\rangle_o+\frac 12{|\xi |}^2(-\cos2\theta +\overline{\coth}_{\sqf}{|\xi |}%
^2).
\label{e77}
\end{eqnarray}

According to (74)$-$(77), SU$_f(1,1)$ squeezing of the states $|\xi
,\q,f{\rangle }_{e(o)}$ occurs as long as
\begin{eqnarray}
&&{\pm }\cos2{\theta }+\overline{\tanh}_{\sqf}{|\xi |}^2<0, \label{e78}\\
&&{\pm }\cos2{\theta }+\overline{\coth}_{\sqf}{|\xi |}^2<0,  \label{e79}
\end{eqnarray}
for one of the sign choices. Choose ${\theta }$ to be $\frac \pi 2$ or $0$,
so that ${\pm }\cos2\theta =-1$. Conditions (78) and (79) may or may not be realized,
depending on the assumed functional form of $f(n)$. When $|f(n)|\leq |f(n+1)|$
($f(n)=1$ and $f(n)=\sqrt{[n]/n}$ included), $\overline{\tanh}_{\sqf}{|\xi |}^2<1$
for $|\xi |\leq f^2(1)$. Thus, condition (78) can be satisfied for $|\xi |\leq f^2(1)$
when $|f(n)|\leq |f(n+1)|$. When $f(n)=\frac1{\sqrt{n^{1-p}}}$ ($p\geq 1$) ($f(n)=1$ included),
from (18), (19), (71) and (72), we have
\begin{equation}
\overline{\coth}_{\sqf}x
= \frac
{\sum\limits_{n=0}^\infty
\frac{x^{2n}}{{[(2n)!(2n+|\q|)!]}^p } }
{\sum\limits_{n=0}^\infty
\frac{x^{2n+1}}{{[(2n+1)!(2n+1+|\q|)!]}^p } }
\equiv \overline{\coth}_{{\q}p}x,
\label{e80}
\end{equation}
where $x={|\xi |}^2$.
In fact, for arbitrary fixed values of $\q$ and $p$, there surely exists some range of $x$
values such that $\overline{\coth}_{{\q}p}x<1$. As shown in Ref. [44], it is indeed true for
$p=1$ (i.e., $f(n)=1$). To make the above statement clear, we plot $\overline{\coth}_{{\q}p}x$
against $x$ for various $\q$ and $p$ in Fig. 1. Thus, for arbitrary fixed values of $\q$ and $p$,
condition (79) can also be satisfied over some limited range of $x$ values
when $f(n)=\frac1{\sqrt{n^{1-p}}}$ ($p\geq 1$). Moreover, condition (79) also holds when $f(n)=\sqrt{[n]/n}$
[64].

{}From the above discussion, we know that the possibility of occurrence of SU$_f(1,1)$ squeezing
for the even and odd nonlinear charge coherent states, depends on the particular form of $f(n)$.
Indeed, the even nonlinear charge coherent states can exhibit SU$_f(1,1)$ squeezing when $|f(n)|\leq |f(n+1)|$;
the odd states can also do when $f(n)=\frac1{\sqrt{n^{1-p}}}$ ($p\geq 1$) and $f(n)=\sqrt{[n]/n}$.
It is mentioned that for the usual even (odd) charge coherent states and the even (odd)
$q$-deformed ones described by $f(n)=1$ and $f(n)=\sqrt{[n]/n}$ respectively,
a detailed discussion of SU(1,1) and SU$_q(1,1)$ squeezing has been presented in Refs. [44,64].

It is easy to verify that the nonlinear charge coherent states satisfy
the equality in (67) and that
${\langle}{({\Delta}X_{1})}^2{\rangle}={%
\langle}{({\Delta}X_{2})}^2{\rangle}$.
Therefore, the nonlinear charge coherent states, contrary to the even and
odd ones, are not SU$_f(1,1)$ squeezed.


\subsection*{\boldmath 5.2. Single-mode $f$-squeezing}

\mbox{}\hspace{6mm}In analogy with the definition of single-mode $q$-squeezing [64,97],
which is a $q$-deformed analouge to single-mode squeezing [27,44,99], we introduce
single-mode $f$-squeezing [82,83] in terms of the Hermitian $f$-deformed quadrature operators
for the individual modes
\begin{eqnarray}
&&Y_1 =\frac{A_1^{\dagger}+A_1}2,\qquad Y_2=\frac{{\rm i}(A_1^{\dagger}-A_1)}2,
\nonumber \\
&&Z_1 =\frac{A_2^{\dagger}+A_2}2,\qquad Z_2=\frac{{\rm i}(A_2^{\dagger}-A_2)}2,
\end{eqnarray}
which satisfy the commutation relations
\begin{equation}
\lbrack Y_1,Y_2]=\frac {\rm i}2[A_1,A_1^{\dagger}],\qquad [Z_1,Z_2]=\frac
{\rm i}2[A_2,A_2^{\dagger}],  \label{e82}
\end{equation}
and the uncertainty relations
\begin{equation}
{\langle }{({\Delta }Y_1)}^2{\rangle }{\langle }{({\Delta }Y_2)}^2{\rangle }{%
\geq }\frac 1{16} {|{\langle }[A_1,A_1^{\dagger}]{\rangle }|}^2,\qquad {%
\langle }{({\Delta }Z_1)}^2{\rangle }{\langle }{({\Delta }Z_2)}^2{\rangle }{%
\geq }\frac 1{16}{|{\langle }[A_2,A_2^{\dagger}]{\rangle }|}^2.  \label{e83}
\end{equation}
A state is said to be single-mode $f$-squeezed if
\begin{equation}
{\langle }{({\Delta }Y_i)}^2{\rangle }<\frac 14 |{\langle }[A_1,A_1^{\dagger}]{%
\rangle }|,\qquad {\langle }{({\Delta }Z_i)}^2{\rangle }<\frac 14 |{%
\langle }[A_2,A_2^{\dagger}]{\rangle }|~\hspace{5mm}(i=1{\rm \ or\ }2).  \label{e84}
\end{equation}
Obviously, single-mode squeezing and $q$-squeezing are the two special cases of
single-mode $f$-squeezing with $f(N_i)=1$ and $f(N_i)=\sqrt{[N_i]/N_i}$, respectively. Therefore,
single-mode $f$-squeezing is a natural extension of single-mode squeezing and $q$-squeezing.

{}For the even and odd nonlinear charge coherent states, it always follows that
\begin{equation}  \label{e85}
{}_{e(o)}{\langle}A_1{\rangle}_{e(o)}={}_{e(o)}{\langle}A_2{\rangle}_{e(o)}=
{}_{e(o)}{%
\langle}A_1^2{\rangle}_{e(o)}={}_{e(o)}{\langle}A_2^2{\rangle}_{e(o)}= {}_{e(o)}{%
\langle}A_1^{\dagger}A_2{\rangle}_{e(o)}=0.
\end{equation}
Thus, the fluctuations are given by
\begin{eqnarray}
{}_{e(o)}{\langle}{({\Delta}Y_{1})}^2{\rangle}_{e(o)}={}_{e(o)}{\langle}{({\Delta%
}Y_{2})}^2{\rangle}_{e(o)}&=& \frac{1}{4}\left\{ {}_{e(o)}{\langle}[A_1,A_1^{\dagger}]{%
\rangle}_{e(o)}+2\, {}_{e(o)}{\langle}A_1^{\dagger}A_1{\rangle}_{e(o)}\right\}  \nonumber
\\
&>&\frac{1}{4}{}_{e(o)}{\langle}[A_1,A_1^{\dagger}]{\rangle}_{e(o)},
\end{eqnarray}
\begin{eqnarray}
{}_{e(o)}{\langle}{({\Delta}Z_{1})}^2{\rangle}_{e(o)}={}_{e(o)}{\langle}{({\Delta%
}Z_{2})}^2{\rangle}_{e(o)}&=& \frac{1}{4}\left\{ {}_{e(o)}{\langle}[A_2,A_2^{\dagger}]{%
\rangle}_{e(o)}+2\, {}_{e(o)}{\langle}A_2^{\dagger}A_2{\rangle}_{e(o)}\right\}  \nonumber
\\
&>&\frac{1}{4}{}_{e(o)}{\langle}[A_2,A_2^{\dagger}]{\rangle}_{e(o)}.
\end{eqnarray}
This shows that there is no single-mode $f$-squeezing in both even and odd nonlinear
charge coherent states. The same situation occurs for the nonlinear charge coherent states [100].


\subsection*{\boldmath 5.3. Two-mode $f$-squeezing}

\mbox{}\hspace{6mm}In analogy with the definition of two-mode $q$-squeezing [64,97],
which is a $q$-deformed analouge to two-mode squeezing [44,99,101], we introduce
two-mode $f$-squeezing [82,83] in terms of the Hermitian $f$-deformed quadrature operators
for the two modes
\begin{equation}  \label{e88}
W_1=\frac{Y_1+Z_1}{\sqrt{2}}=\frac{1}{\sqrt{8}}(A_1^{\dagger}+A_2^{\dagger}+A_1+A_2),\quad
W_2=\frac{Y_2+Z_2}{\sqrt{2}}=\frac{\rm i}{\sqrt{8}}(A_1^{\dagger}+A_2^{\dagger}-A_1-A_2),
\end{equation}
which satisfy the commutation relation
\begin{equation}  \label{e83}
[W_1,W_2]=\frac{1}{4}{\rm i}\left\{[A_1,A_1^{\dagger}]+[A_2,A_2^{\dagger}]\right\}
\end{equation}
and the uncertainty relation
\begin{equation}  \label{e84}
{\langle}{({\Delta}W_{1})}^2{\rangle}{\langle}{({\Delta}W_{2})}^2{\rangle}{%
\geq}\frac{1}{64} {|{\langle }[A_1,A_1^{\dagger}]{\rangle }+{\langle }[A_2,A_2^{\dagger}]{%
\rangle } |}^2.
\end{equation}
A state is said to be two-mode $q$-squeezed if
\begin{equation}  \label{e85}
{\langle}{({\Delta}W_{i})}^2{\rangle}<\frac{1}{8} {|{\langle }[A_1,A_1^{\dagger}]{%
\rangle }+{\langle }[A_2,A_2^{\dagger}]{\rangle } |} ~ \hspace{8mm}(i=1 {\rm \ or\ }2).
\end{equation}
Obviously, two-mode squeezing and $q$-squeezing are the two special cases of
two-mode $f$-squeezing with $f(N_i)=1$ and $f(N_i)=\sqrt{[N_i]/N_i}$, respectively. Therefore,
two-mode $f$-squeezing is a natural extension of two-mode squeezing and $q$-squeezing.

{}For the even and odd nonlinear charge coherent states, the fluctuations
are given by
\begin{eqnarray}
{}_{e(o)}{\langle}{({\Delta }W_1)}^2{\rangle}_{e(o)}&=&{}_{e(o)}{\langle}{({%
\Delta }W_2)}^2{\rangle}_{e(o)}=\frac 12\left\{{}_{e(o)}{\langle}{({\Delta }Y_1)%
}^2{\rangle}_{e(o)}+{}_{e(o)}{\langle}{({\Delta }Z_1)}^2{\rangle}_{e(o)}\right\}
\nonumber \\
&=&\frac 12\left\{{}_{e(o)}{\langle }{({\Delta }Y_2)}^2{\rangle }_{e(o)}+{}_{e(o)}%
{\langle }{({\Delta }Z_2)}^2{\rangle }_{e(o)}\right\}  \nonumber \\
&=&\frac 18 \Bigl\{ {}_{e(o)}{\langle}[A_1,A_1^{\dagger}]{\rangle}_{e(o)}+ {}_{e(o)}{%
\langle}[A_2,A_2^{\dagger}]{\rangle}_{e(o)} +2\,{}_{e(o)}{\langle}A_1^{\dagger}A_1{\rangle}%
_{e(o)} \nonumber \\
&&\mbox{} +2\,{}_{e(o)}{\langle}A_2^{\dagger}A_2{\rangle}_{e(o)} \Bigr\}  \nonumber\\
&>&\frac{1}{8} \left\{ {}_{e(o)}{\langle}[A_1,A_1^{\dagger}]{\rangle}_{e(o)}+ {}_{e(o)}{%
\langle}[A_2,A_2^{\dagger}]{\rangle}_{e(o)} \right\}.
\end{eqnarray}
This shows that there is no two-mode $f$-squeezing in both even and odd nonlinear
charge coherent states. On the contrary, there is such $f$-squeezing in
the nonlinear charge coherent states, depending on the particular form of $f(n)$.
As shown in Refs. [44,64,100], two-mode $f$-squeezing does exist for the nonlinear
charge coherent states with $f(n)=\frac1{\sqrt{n^{1-p}}}$ ($p\geq 1$) and $f(n)=\sqrt{[n]/n}$.


\subsection*{\boldmath 5.4. Two-mode $f$-antibunching}

\mbox{}\hspace{6mm}In analogy with the definition of two-mode $q$-antibunching [64,97],
which is a $q$-deformed analouge to two-mode antibunching [44,99], we introduce
a two-mode $f$-correlation function as [82]
\begin{equation}
g^{(2)}(0)\equiv \frac{{\langle }{(A_1^{\dagger}A_2^{\dagger})}^2{(A_1A_2)}^2{\rangle }}{{{%
\langle }A_1^{\dagger}A_2^{\dagger}A_1A_2{\rangle }}^2}=\frac{{\langle }:{(N_1f^2(N_1)N_2f^2(N_2))}%
^2:{\rangle }}{{{\langle }N_1f^2(N_1)N_2f^2(N_2){\rangle }}^2}
=\frac{ {\langle }{(K_{-}^{\dagger})}^2K_{-}^2{\rangle }  } { { {\langle }
K_{-}^{\dagger}K_{-} {\rangle } }^2 }, \label{e93}
\end{equation}
where $A_i$ and $A_i^{\dagger}$ represent the annihilation and creation operators
of $f$-deformed photons of a nonlinear light field for the $i${\sl th} mode and $:\,:$ denotes normal
ordering. Physically, $g^{(2)}(0)$ is a measure of $f$-deformed two-photon
correlations in the nonlinear two-mode field and is related to the $f$-deformed
two-photon number distributions. A state is said to be two-mode $f$-antibunched if
\begin{equation}
g^{(2)}(0)<1.  \label{e94}
\end{equation}
Obviously, two-mode antibunching and $q$-antibunching are the two special cases of
two-mode $f$-antibunching with $f(N_i)=1$ and $f(N_i)=\sqrt{[N_i]/N_i}$, respectively. Therefore,
two-mode $f$-antibunching is a natural extension of two-mode antibunching and $q$-antibunching.

{}For the even and odd nonlinear charge coherent states, we have
\begin{eqnarray}
&&g_e^{(2)}(0) =  \overline{\coth}_{\sqf}^2 {|\xi |}^2, \\
&&g_o^{(2)}(0) = \overline{\tanh}_{\sqf}^2{|\xi |}^2.
\end{eqnarray}
{}From the above discussion about the function $\overline{\coth}_{\sqf}{|\xi |}%
^2$ $(\overline{\tanh}_{\sqf}{|\xi |}^2)$, we see that depending on the particular form of $f(n)$,
$g_{e(o)}^{(2)}(0)$ can be less than $1$
over some limited range of $|\xi|$ values, producing two-mode $f$-antibunching.
Indeed, the even nonlinear charge coherent states can display such $f$-antibunching
when $f(n)=\frac1{\sqrt{n^{1-p}}}$ ($p\geq 1$) and $f(n)=\sqrt{[n]/n}$; the odd states can also do
when $|f(n)|\leq |f(n+1)|$.  On the contrary, for the nonlinear charge coherent states
we have $g^{(2)}(0)=1$ so that no two-mode $f$-antibunching exists.

It can be shown that in the two special cases of $f(n)=1$ and $f(n)=\sqrt{[n]/n}$, the nonclassical
properties of the usual even (odd) charge coherent states studied in Ref. [44] and the even (odd)
$q$-deformed ones in Ref. [64], are  retrieved as expected, respectively.



\section*{6. Summary}

\mbox{}\hspace{6mm}Let us sum up the results obtained in the present paper.

(1) The (over)completeness has been proved of the even and odd nonlinear charge coherent states,
defined as two orthonormalized eigenstates of both the square of the pair $f$-deformed annihilation
operator $a_1f(N_1)a_2f(N_2)$ and the charge operator $Q=N_1-N_2$.

(2) The even (odd) nonlinear charge coherent states have been shown to be
generated by a suitable average over the $U(1)$-group (caused by the charge operator) action
on the product of nonlinear coherent states and even (odd) nonlinear coherent states.
They have also been demonstrated to be generalized entangled nonlinear coherent states.

(3) The $D$-algebra of the SU$_f(1,1)$ generators corresponding to the even and odd
nonlinear charge coherent states has been realized in a differential-operator matrix form.

(4) It has been shown that the even (odd) nonlinear charge coherent states have the possibility
of existence of the nonclassical properties in the particular form of $f(n)$, and
exhibit SU$_f(1,1)$ squeezing for $|f(n)|\leq |f(n+1)|$ ($f(n)=\frac1{\sqrt{n^{1-p}}}$ ($p\geq 1$))
and two-mode $f$-antibunching for $f(n)=\frac1{\sqrt{n^{1-p}}}$ ($p\geq 1$) ($|f(n)|\leq |f(n+1)|$),
but neither single-mode nor two-mode $f$-squeezing.

\section*{Acknowledgments}

\mbox{}\hspace{6mm}X.-M.L.\ would like to acknowledge Master Y. Shi for
discussions in drawing the figures. We also thank postgraduate S. Zhu
for her help in printing the manuscript. This work was supported by
the National Natural Science Foundation of China under Grant No. 11075014.


\newpage

\section*{Figure caption}

\mbox{}\hspace{6mm}Fig. 1. $\overline{\coth}_{{\q}p}x$ against $x$ for
$p=2,~3,~4$,
with (a) $\q=0$, (b) $\q={\pm}1$ and (c) $\q={\pm}2$.



\newpage
\begin{figure}
\centering
\includegraphics[width=1.00\textwidth]{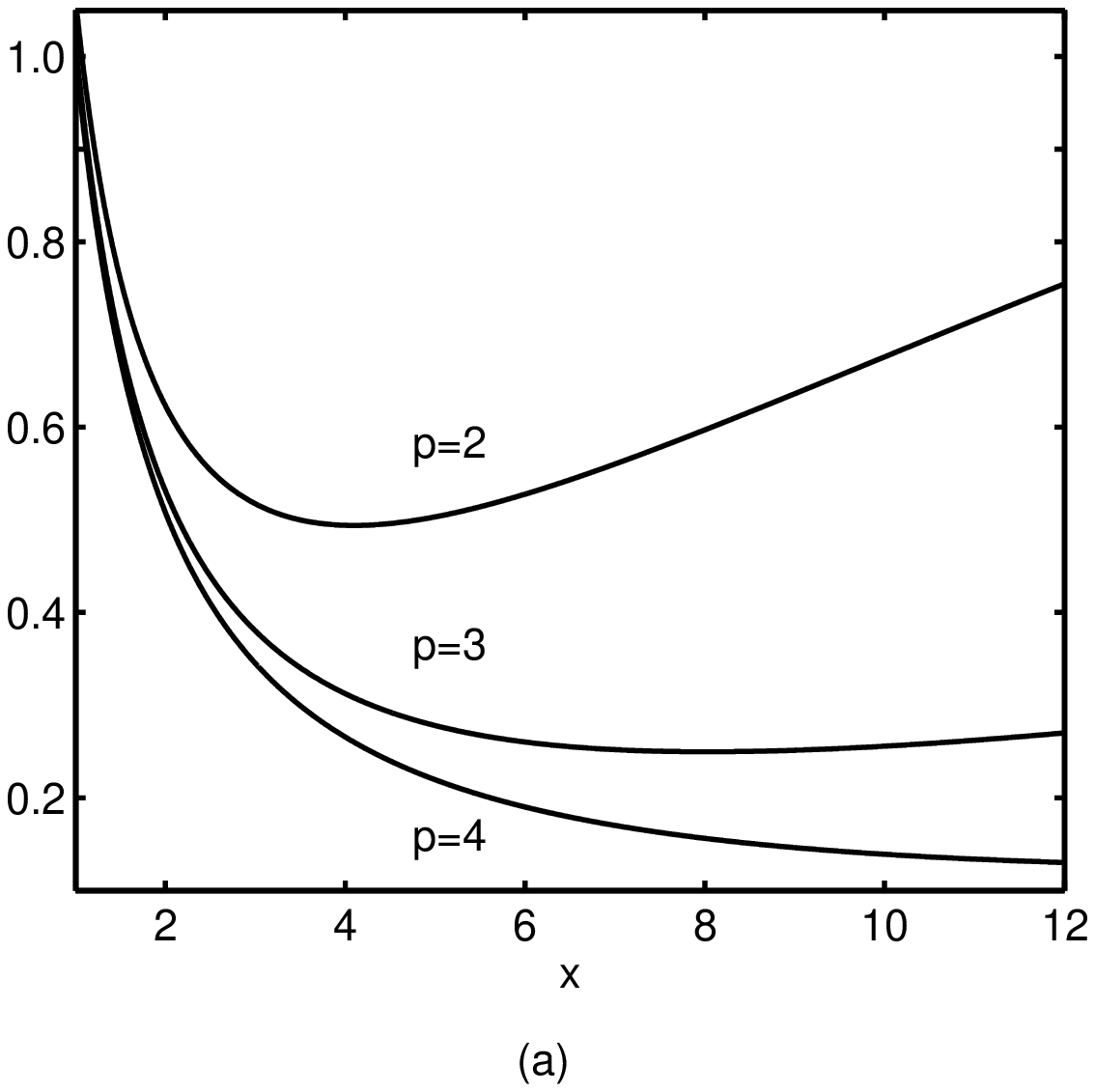}
\end{figure}

\newpage
\begin{figure}
\centering
\includegraphics[width=1.00\textwidth]{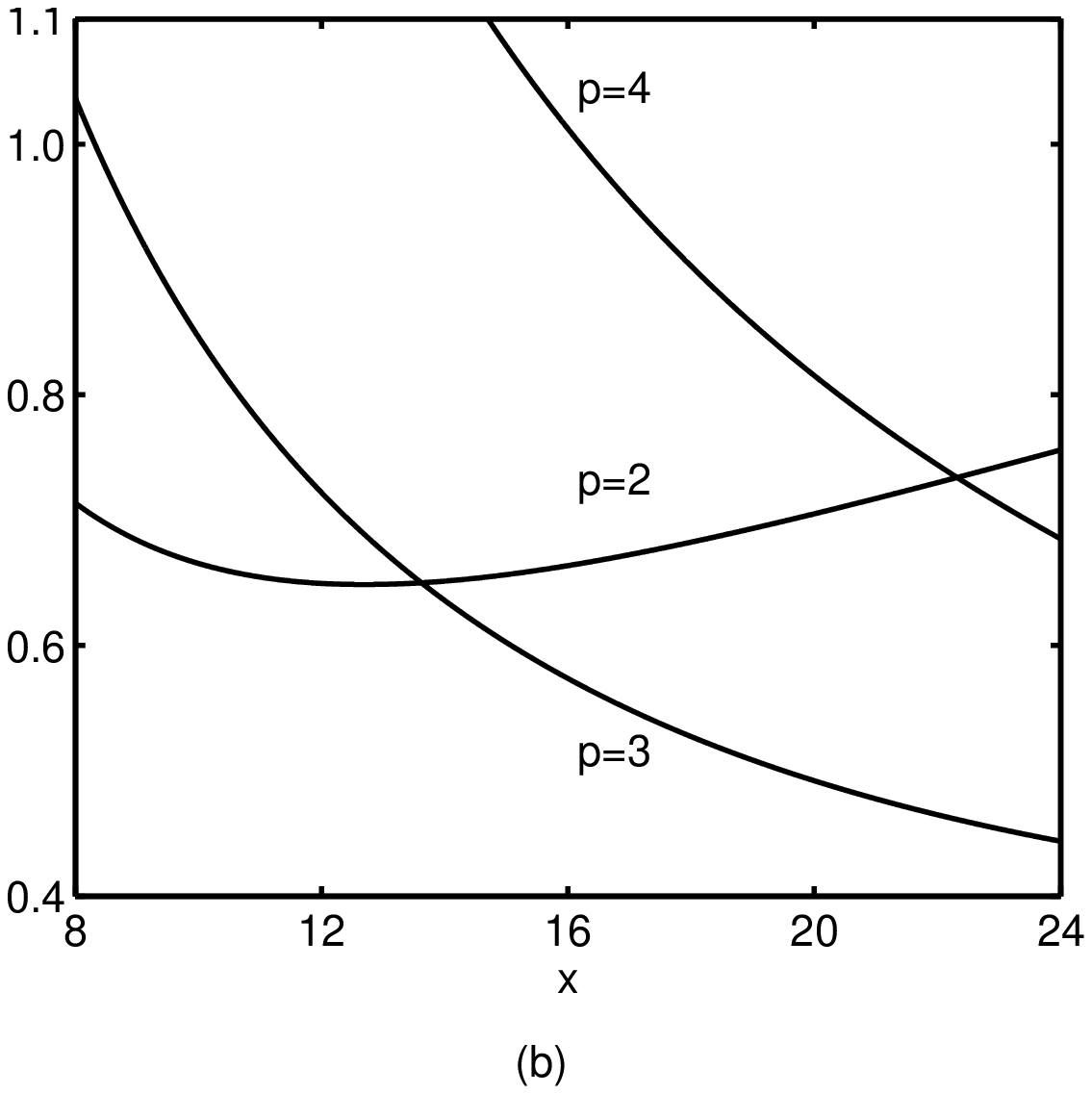}
\end{figure}

\newpage
\begin{figure}
\centering
\includegraphics[width=1.00\textwidth]{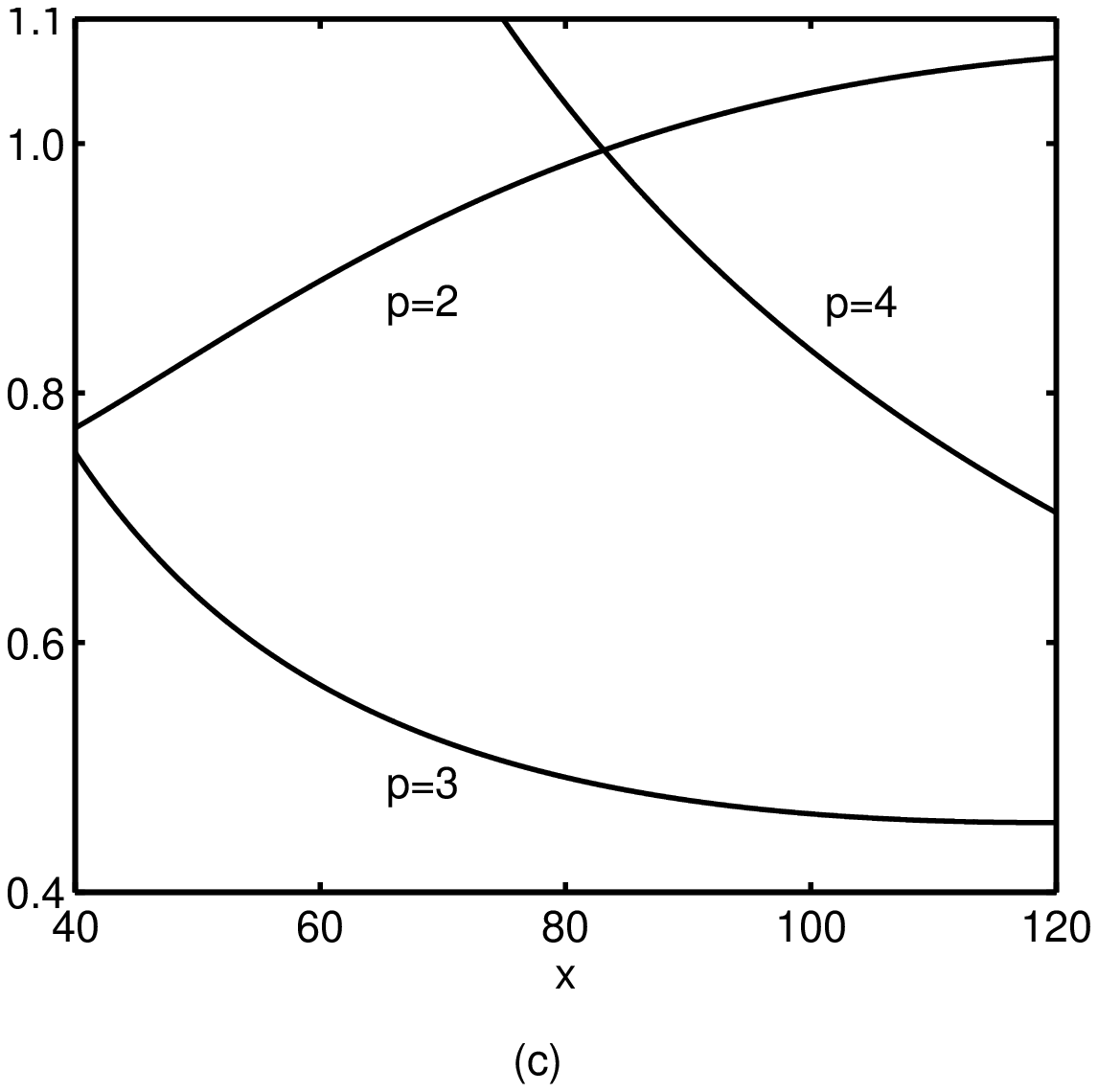}
\end{figure}

\end{document}